# Fast Characterization of Dispersion and Dispersion Slope of Optical Fiber Links using Spectral Interferometry with Frequency Combs

V. R. Supradeepa, *Student Member, IEEE,* Christopher M. Long, Daniel E. Leaird, *Senior Member, IEEE,* Andrew M. Weiner, *Fellow, IEEE*

*Abstract*— We demonstrate fast characterization (~1.4 μs) of both the dispersion and dispersion slope of long optical fiber links (~25 km) using dual quadrature spectral interferometry with an optical frequency comb. Compared to previous spectral interferometry experiments limited to fiber lengths of meters, the long coherence length and the periodic delay properties of frequency combs, coupled with fast data acquisition, enable spectral interferometric characterization of fibers longer by several orders of magnitude. We expect that our method will be useful to recently proposed lightwave techniques like coherent WDM and to coherent modulation formats by providing a real time monitoring capability for the link dispersion. Another area of application would be in stabilization of systems which perform frequency and timing distribution over long fiber links using stabilized optical frequency combs.

*Index Terms*— Optical fiber dispersion, Optical fiber measurements, Ultrafast pulse measurement, Frequency combs, Spectral interferometry.

## I. INTRODUCTION

In recent years there has been significant work on novel paradigms in optical communications where the phase of the optical carriers is controlled to obtain better spectral efficiency with reduced crosstalk [1-2]. There also has been significant renewal of interest in coherent optical communications [3] where again the phase plays an important role. In practical systems involving installed fiber, the link dispersion is not expected to remain fixed but vary slightly over time. This will also be the case with reconfigurable optical networks where the data might be carried by different links, each with different dispersion properties. Dispersion along these links will not only manifest itself as temporal distortion but also as frequency dependent phase variations of the transmitted signal. Therefore a real time dispersion monitoring tool which can perform fast measurements with low power requirements is desirable. Another area where this would be useful is in timing and frequency transfer over fiber links [4-6]. Fiber delay fluctuations which occur in submillisecond time scales manifest themselves as phase noise in the transmitted signal. In situations where broad bandwidths (i.e. short pulses) are used, dispersion variations which cause frequency dependent delay fluctuations would also have to be compensated faster than the fluctuation time scales.

Measuring dispersion properties in fiber is well established [7]. However, common shortcomings are relatively long acquisition times and difficulty in precisely measuring large dispersions expected in long fiber links. In temporal domain measurements, large dispersion manifests itself as large pulse broadening where exact determination of dispersion becomes difficult due to varying pulse shapes. Other methods like the radio-frequency phase shift technique [7] require wavelength scanning and are hence slow.

In this work we perform fast measurement(<1.4 μs, limited only by camera integration time) of both dispersion and dispersion slope directly by measuring the $2^{nd}$ and $3^{rd}$ order spectral phase of a high repetition rate optical frequency comb after dispersive propagation over ~25 km length of fiber. Two primary factors contribute to the success of our experiment; one is the use of high repetition rate frequency combs and the second is going to a zero delay scheme of spectral interferometry. Firstly, conventional spectral interferometry has been used previously for dispersion measurement but with using white light sources (see [10] for example); however the need to resolve interferometric fringes associated with nonzero delay constrained the fiber length to a few metres. In our experiments we exploit the periodic nature of delay and the long coherence lengths associated with optical frequency combs to perform direct interferometric measurements of spectral phase for lengths up to 50 km of standard SMF [9], many orders of magnitude longer than previous spectral interferometry experiments. Secondly, in conventional implementations of spectral interferometry, even with a frequency comb removing the effect of delay based fringes, phase retrieval becomes difficult owing to the requirement of a large delay between interfering pulses. This significantly increases resolution requirements and for dispersed high repetition rate pulses (>50% duty factor) becomes impossible. So, our measurement uses a zero delay version of spectral interferometry known as dual-quadrature spectral interferometry. Here by using polarization demultiplexing to encode the complete interference signal

Manuscript received July 23, 2009. This work was supported by the DARPA/ARO under the Grant W911NF-07-1-0625 and by the NSF under the Grant ECCS-0601692.

All the authors are with the School of Electrical and Computer Engineering, Purdue University, West Lafayette, IN 47907, USA. (Email: supradeepa@purdue.edu).

(both in-phase and quadrature components) the need for a substantial delay for unambiguous phase retrieval is eliminated. This also allows for reduced resolution and complexity requirements [8, 9].

## II. EXPERIMENTAL SETUP

Fig. 1 shows the experimental setup. The input frequency comb is generated using an optical frequency comb generator which generates periodic sidebands of a driving CW laser through a phase modulator inside a cavity [11]. The bandwidth of the comb is approximately 500 GHz, with spectral lines spaced by 10 GHz centred at 1542 nm. The input passes through a fiber splitter, and one part of it is sent through the fiber link, the output of which we will refer to as the signal. The other part of the input is the reference and used to extract the spectral phase of the transmitted signal. Since this method measures the spectral phase in a relative fashion, the original spectral phase of the input is not important.

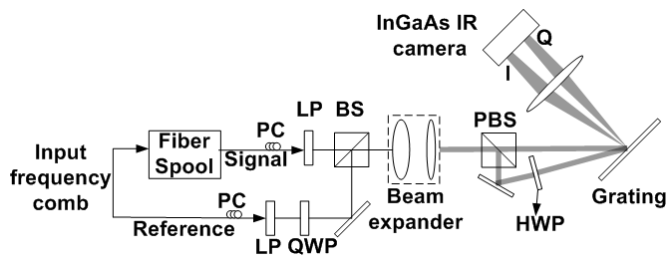

Fig. 1. Experimental Setup; LP – Linear polarizer, QWP – Quarter wave plate, BS – Beam splitter, PBS – Polarizing beam splitter, HWP – Half wave plate, PC – Polarization controller.

Using commercially available optical collimators (OFR, TA-18), the signal and reference are sent into free space. The signal to be measured is linearly polarized at a 45° angle while the reference is circularly polarized. These two beams are combined followed by a high resolution spectrometer consisting of a 10X beam expander, a grating with 1100 line/mm, and an InGaAs IR camera. The pixel dimension of the camera in the dispersion direction is 25 μm, and there are 512 pixels. The spectrometer resolution is 3.33 GHz per pixel which gives a line-to-line spacing of 3 pixels on the camera (corresponding to the frequency comb spacing of 10 GHz). The spectrometer simultaneously measures the interferograms in both polarizations (corresponding to the in-phase and quadrature terms) by mapping them to different physical locations on the camera. For both channels the measured spectrometer crosstalk between adjacent comb lines is ~5%. We retrieve waveform information unambiguously from a single frame of camera data with 1.4-μs integration time. Other than the ability to make fast measurements, a fast acquisition is also needed because fiber length fluctuations can occur in sub ms time scales and this can affect measurements if the acquisition time is large. In our case we expect the 1.4-μs integration time to be orders of magnitude faster than the fluctuation time scales allowing for clean measurements. Regarding power requirements, the signal and reference power levels for the data shown in this paper were ~0.1mW and ~0.2mW, respectively. However, we have verified similar performance (looking at error residuals) even for ten times lower powers. A mathematical description and retrieval information is discussed in [9]. A point to note here is that there is no fundamental limitation on how much signal bandwidth can be handled. By choosing a camera with more pixels, the bandwidth can be significantly increased depending on the requirements of the application.

## III. RESULTS

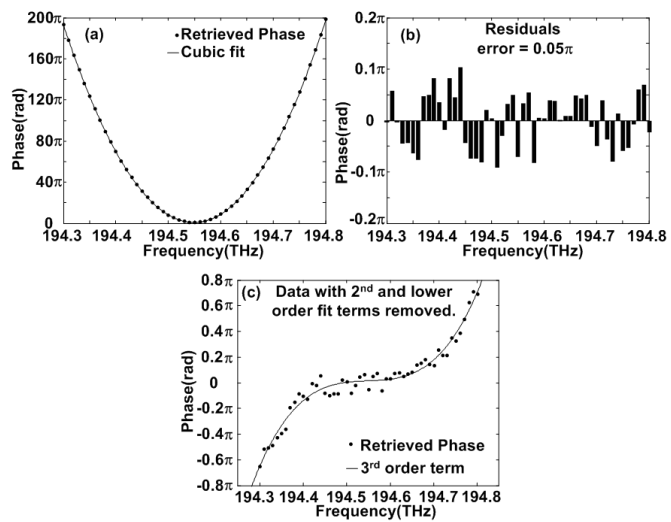

Fig. 2. 25km standard SMF, (a) Retrieved spectral phase and cubic fit; (b) Residuals of the fit; (c) Spectral phase plot after removing the quadratic phase component (dispersion).

The first set of experiments was done with a 25-km spool of standard SMF (OFS-Fitel). Fig 2(a) shows the retrieved spectral phase and a cubic fit to it. Fig 2(b) shows the residuals between the fit and the retrieved spectral phase. The error, defined as the standard deviation of the residuals, is $0.05\pi$, indicating a very good fit. The choice of using a cubic function was made because at this bandwidth both the cubic term (which is related to the dispersion slope) and the quadratic term (which gives the dispersion) are expected to be significant. This was confirmed by significantly higher errors (more than twice) obtained while fitting a quadratic only curve to the retrieved phase. However since the quadratic contribution dominates the cubic contribution, the curve still looks largely like a parabola. To show the presence of the cubic we subtracted the quadratic term and plot the difference in Fig 2(c). We see a clear cubic behavior from this figure. From the parameters of the fit we estimate the dispersion to be 393.6 ps/nm and the dispersion slope to be 1.50 ps/nm^2.

The next experiment is done to check the dispersion properties of a dispersion compensating fiber (DCF) module that is designed to compensate the 25-km link (OFS-Fitel, module type – EWBDK: 404). Fig. 3(a) shows the retrieved spectral phase and the cubic fit. From the parameters of the fit we obtain the total dispersion as -395.3 ps/nm and dispersion slope -1.39ps/nm^2. The data provided by the vendor for the DCF fiber module is a dispersion of -394ps/nm and a dispersion slope is -1.36ps/nm^2 at 1542nm. Our results and the numbers provided by the vendor show very good agreement.

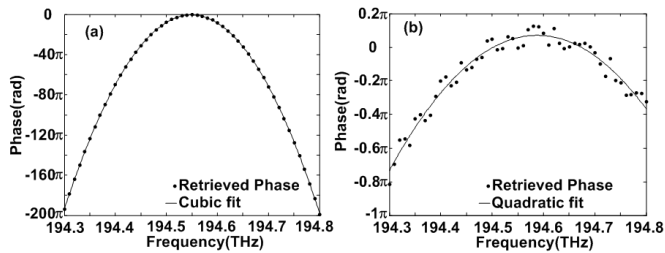

Fig. 3 (a) Retrieved spectral phase for a DCF module for the 25km SMF link with a cubic fit; (b) Retrieved spectral phase for the dispersion compensated link with a quadratic fit.

Although the vendor specifications for the SMF module say that it is "matched in dispersion to the DCF module," our measured dispersion values show slight differences between the SMF and DCF. To further investigate, we performed an experiment with the concatenation of the 25-km SMF link and the DCF module. Adding the dispersion values obtained from our measurements of the SMF and DCF taken one at a time, we expect a small net dispersion of -1.7 ps/nm and a net slope of 0.11 ps/nm^2. The retrieved spectral phase of the concatenated link is shown in Fig 3(b). The quadratic fit to the data gives a net dispersion of -1.5 ps/nm, very close to what we expected, while the cubic phase term is negligible (below our measurement sensitivity). These experiments are indicative of the high precision in our technique allowing us to measure small dispersion (better than ~1.5ps/nm) while also simultaneously being able to measure very large dispersion (~395ps/nm), thus providing a large dynamic range of operation. As a further check, these results are in qualitative agreement with experiments previously done in our group using the same fiber modules for dispersion compensated transmission of short pulses over fiber lengths up to 50 km [12].

Before we conclude we will briefly comment on issues pertaining to periodic sampling of spectral phase when using frequency combs. Looking at the unwrapped retrieved phase, we see that the phase difference between adjacent data points increases away from the center frequency and is several $2\pi$ towards the edges of the spectrum. This prompts discussion on the uniqueness of the unwrapping process. Since all the data points contribute together to the unwrapping, for a phase profile which is primarily quadratic, the unwrapping is unique as long as the minimum phase difference between adjacent data points (in our data observed near center frequency) is less than $\pi$. For our ~10(9.952) GHz comb spacing centered at 1542nm, unique unwrapping is possible for dispersion up to ~1275 ps/nm (corresponds to uncompensated SMF distance between zero and ~77km). For larger dispersion the measurement yields the actual dispersion modulo ~1275 ps/nm. Note that lower repetition rate frequency combs allow for unambiguous measurement over even larger range of dispersion (the total unambiguous dispersion range scales as the square of inverse of the repetition rate). A more detailed discussion on this issue is presented in [9].

## IV. SUMMARY AND CONCLUSIONS

We have demonstrated a method to perform fast (~1.4 μs, limited by camera integration time) characterization of dispersion and dispersion slope of long fiber links by performing interferometric spectral phase measurements with frequency combs. In this work we measured the dispersion properties over a 500 GHz bandwidth of a 25 km standard SMF link, a matched DCF module, and the link combining the two. This method is easily scalable to wider bandwidths as required by the application. The fast acquisition time of our system (significantly faster than fiber fluctuations), coupled with state of the art frequency combs which can have linewidths even below 1 Hz [13], provides potential to extend these measurements to fiber lengths of hundreds or even thousands of kms. This method has promise as a real time dispersion monitoring tool for next generation optical communications systems where dispersion monitoring will be even more valuable owing to increased bandwidths and phase sensitive formats.